\documentclass[12pt]{article}

\usepackage{epsf}
\def\lsim{\raise0.3ex\hbox{$\;<$\kern-0.75em\raise-1.1ex\hbox{$\sim\;$}}}
\def\gsim{\raise0.3ex\hbox{$\;>$\kern-0.75em\raise-1.1ex\hbox{$\sim\;$}}}

\def\ord#1{{\cal O}(#1)}
\def\lsim{\raise0.3ex\hbox{$\;<$\kern-0.75em\raise-1.1ex\hbox{$\sim\;$}}}
\def\gsim{\raise0.3ex\hbox{$\;>$\kern-0.75em\raise-1.1ex\hbox{$\sim\;$}}}
\begin{document}


\pagestyle{empty}
\vspace{-0.6in}
\begin{flushright}
ROMA-1311/01\\
FERMILAB-Pub-01/048-T 
\end{flushright}
\vskip 1.2 cm                                                                   
\centerline{\Large{\bf{Signals of Supersymmetric Flavour Models}}}
\centerline{\Large{\bf{ in 
 \boldmath$B$ Physics}}}
\vspace{1cm}  
\centerline{\bf A. Masiero, M. Piai} 
\vspace{0.3cm}
\centerline{SISSA -- ISAS, Via
  Beirut 4, I-34013, Trieste, Italy and} 
\centerline{INFN, Sezione di
  Trieste, Trieste, Italy.}  
\vspace{0.5cm}
\centerline{\bf A. Romanino } 
\vspace{0.3cm}
\centerline{Fermi National Accelerator Laboratory}
\centerline{P.O. Box 500, Batavia, IL 60510, USA}
\vspace{0.5cm}
\centerline{\bf L. Silvestrini} 
\vspace{0.3cm}
\centerline{INFN, sez. di Roma and Dip. di
  Fisica,} 
\centerline{Univ. di Roma ``La Sapienza'', P.le A. Moro,
  I-00185 Roma, Italy.}  

\begin{abstract}
  If the mechanism of Supersymmetry breaking is not flavour blind,
  some flavour symmetry is likely to be needed to prevent excessive
  flavour changing neutral current effects.  We discuss two flavour
  models (based respectively on a $U(2)$ and on a $SU(3)$ horizontal
  symmetry) providing a good fit to fermion masses and mixings and
  particularly constraining the supersymmetry soft breaking terms.  We show
  that, while reproducing successfully the Standard Model fit of the
  unitarity triangle, it is possible to obtain sizable deviations from
  the Standard Model predictions for three very clean $B$-physics
  observables: the time dependent CP asymmetries in $B_d \rightarrow
  J/\psi K^0$ and in $B_s \rightarrow J/\psi \phi$ and the
  $B_s-\bar{B}_s$ mass difference.  Our analysis exhibits with two
  explicit realizations that in supersymmetric theories with a new
  flavour structure in addition to the Yukawa matrices there exist
  concrete potentialities for revealing supersymmetry indirectly in
  theoretically clean $B$-physics observables.
\end{abstract}
\vfill\eject
\pagestyle{empty}\clearpage
\setcounter{page}{1}
\pagestyle{plain}
\newpage
\pagestyle{plain} \setcounter{page}{1}                                          
\section{Introduction}

For the last two decades, the indirect search for Supersymmetric
(SUSY) signals through Flavour Changing Neutral Current (FCNC) and CP
violating processes has proven to be a crucial complementary tool to
direct accelerator search~\cite{review}. After the end of the LEP era,
our hopes for a detection of SUSY particles focus on the upgraded
Tevatron and even more on LHC, the resolutive machine for low-energy
SUSY. In the years before LHC, the challenge for SUSY hints mostly
relies upon virtual effects in FCNC and CP rare processes. After the
intensive experimental and theoretical work on Kaon physics, and
waiting for the important results on rare $K$ decays, the next
frontier is represented by $B$ physics. Although all of us hope in
some dramatic effect signaling the presence of new physics (for
instance, had the CP asymmetry in $B \to J/\psi K$ ($a_{J/\psi K}$)
settled at the level of $10\%$ there would be no
doubt~\cite{lowasymmetry}), it is likely that we will have to face a
more complicate situation where the information on new physics will be
entangled with the hadronic uncertainties plaguing nonleptonic $B$
decays. In view of this fact, processes as $B_s$--$\bar B_s$ mixing
acquire a crucial relevance in increasing the redundancy of the
Unitarity Triangle (UT) determination, hence allowing for a possible
discrimination among different SUSY extensions of the SM. In this
respect, it proves quite useful to test various classes of low-energy
SUSY models considering, in addition to the stringent constraints from
$K$ physics, the joint information from the mixing and the CP
asymmetries in $B$ physics.

On the other hand, just the severity of the present 
FCNC constraints~\cite{constraints, Bagger, Ciuchini:1998ix}
seems to point to two definite directions: either the mechanism of
SUSY breaking is flavour blind, resulting in the so-called Minimal
Supersymmetric Standard Model (MSSM) with Minimal Flavour Violation (MFV),
or we need some mechanism (based on flavour 
symmetries, alignment, or heavy first generations sfermions, for instance)
 to forbid disastrously large SUSY
contributions to FCNC and CP violating processes arising 
from the new flavour structure of the model.
 As for the former
option, already several detailed analyses of the impact of these
models on FCNC and CP violation have been performed~\cite{MFVanalyses}.
 Concerning the
second possibility, much interesting work has recently focused on the
construction of successful non-abelian flavour 
models~\cite{flavour}-\cite{RVuc},
 mainly concentrating on the
prediction of fermion masses and mixing angles. However, with a few
valuable exceptions, most of these works have not thoroughly
investigated the impact of SUSY contributions to FCNC in relation with
the UT determination. Such attitude was fully justified when the main
objective was the prohibition of too large SUSY effects, but nowadays,
since our goal is the detailed comparison of the SM and SUSY
predictions on FCNC, it is mandatory to reconsider SUSY flavour models
taking into account the specific SUSY contributions to rare processes.

As a first step in this direction, in this Letter we consider two
promising models with nonabelian flavour symmetries, which
particularly constrain the flavour structure of the SUSY soft
breaking terms. We show that it is possible to successfully reproduce
the SM fit of the UT while allowing for sizable deviations from the SM
predictions for three interesting $B$ physics observables: $a_{J/\psi
  K}$, $a_{J/\psi
  \phi}$, the time-dependent CP asymmetry in $B_s \to J/\psi \phi$
decays,  
and $\Delta m_{B_s}$, the $B_s$--$\bar B_s$ mass difference. Our analysis
shows the importance of using theoretically clean $B$-physics
observables in disentangling the SUSY effects in models with viable
flavour structures.

\section{A model with a $U(2)$ flavour symmetry}
\label{sec:U2}

Let us first consider a model based on a $U(2)$ symmetry acting on the
two lighter families \cite{flavour}-\cite{BHR}.  The pattern of fermion
masses and mixing reveals an approximately symmetric structure under
$U(2)$. This symmetry, in fact, suppresses (forbids, in the unbroken
limit) the Yukawa couplings of the two lighter families and the
non-degeneracy of their supersymmetric partners. Moreover, the $U(2)$
symmetry can be considered the residual symmetry unbroken after the
large breaking of an $U(3)$ symmetry by the top Yukawa coupling.  The
fermions of the third, $\psi_3$, and of the first two families,
$\psi_a$, $a=1,2$, have the obvious transformation properties under
$U(2)$. The Higgs fields are assumed to be singlets.  The Yukawa
couplings involving the lighter families are associated to VEVs of SM
singlets breaking the flavour symmetry and coupling to the SM fermions
through non-renormalizable Yukawa interactions. Such VEVs can only
transform as an antidoublet $\phi^a$, an antisymmetric tensor $A^{ab}$
or a symmetric tensor $S^{ab}$ under the flavour
symmetry\footnote{Upper and lower indexes correspond to conjugated
  transformations.}. The two step breaking of the rank two group
$U(2)$ can be accomplished by using only the first two of those
representations: $\phi^a$ and $A^{ab}$. No assumption needs to be made
on the orientation of the corresponding VEVs in the flavour space,
since every choice is equivalent to $\langle \phi \rangle=(0,V)^T$,
$V>0$, $\langle A^{ab} \rangle = v\epsilon^{ab}$, $v>0$ up to an
$U(2)$ transformation.  Notice that $\langle\phi\rangle$ leaves a
residual $U(1)$ unbroken, which protects the mass of the lightest
family. The asymmetric VEV $\langle A^{ab} \rangle$ then breaks the
residual $U(1)$ and gives mass to the lightest family. The interfamily
mass hierarchy is obtained if $V>v$, so that
\begin{equation}
\label{breaking}
U(2) \stackrel{V}{\rightarrow} U(1) \stackrel{v}{\rightarrow} 1 \; .
\end{equation}

Within this framework, we now briefly describe a new model which is a
variation of~\cite{BH2}, to which we refer for a more detailed
discussion of the general framework, and represents an example of how
our understanding of flavour and CP-violation can be affected by new
physics.    We assume that the $U(2)$ breaking is communicated to the
SM fermions $\psi_3$, $\psi_a$ through a Froggatt-Nielsen (FN)
mechanism by an heavy $U(2)$ doublet $\chi^a$ in the same gauge
representation as a whole fermion family.  Since we want $\chi^a$ to
be heavy in the $U(2)$ symmetric limit, we include the conjugated
fields $\bar\chi_a$ in the messenger sector. We work in the context of
a supersymmetric $SU(5)$ GUT. Once $U(2)$ is broken, the light (in the
$U(2)$-symmetric limit) families $\psi_a$ and the heavy copies
$\chi^a$ mix, thus giving rise to the light Yukawa couplings. We also
take into account the possibility that the two SU(5) multiplets $H_1$,
$H_2$ containing the up and down light Higgses mix with heavy copies
$H^{\prime}_1$, $H^{\prime}_2$, $U(2)$ singlets too\footnote{If part
  of the hierarchy $m_b \ll m_t$ is due to an hierarchy between the
  corresponding Yukawa couplings, the latter can be accounted for by a
  mixing in the Higgs sector.}.

Let us now discuss the size of mass terms and VEVs. One simple
possibility is to assume that the mass $M$ of the heavy doublets
$\chi^a$, $\bar\chi_a$ is generated above the $SU(5)$ breaking scale,
$M>M_{\rm GUT}$, and is therefore $SU(5)$ invariant. A small ratio
$V/M$ is then generated if the $U(2)$-breaking takes place at the
$SU(5)$ breaking scale, $V\sim M_{\rm GUT}$. $SU(5)$ breaking
corrections to the heavy mass $M$ will also be correspondingly smaller
than $M$. As for the mass $M^{\prime}$ of the heavy multiplets possibly mixing
with the Higgs multiplets, we will assume it to be of the order of the
GUT scale. The $U(2)$-singlet, $SU(5)$ fiveplet messengers $H^{\prime}_1$,
$H^{\prime}_2$ will therefore be lighter than the $U(2)$ doublet messengers
$\chi^a$, $\bar\chi_a$.  This at the same time accounts for the
empirical relation $m_s/m_b\sim|V_{cb}|$ and for the hierarchy
$m_c/m_t\ll m_s/m_b$, enhances the supersymmetric contributions to
$B$-mixing, improves the agreement of the measured value of
$|V_{ub}/V_{cb}|$ with the prediction of the model in terms of light
quark masses~\cite{RVuc} and might be related to the large mixing in
the neutrino sector indicated by the atmospheric neutrino
anomaly~\cite{RVuc}. Finally, the breaking of the residual $U(1)$
occurs below the GUT scale, $v< M_{\rm GUT}$. As for the
transformation properties of the VEVs $A^{ab}$, $\phi^a$ under
$SU(5)$, the only crucial assumption is that $A^{ab}$ is $SU(5)$
invariant, which accounts for the hierarchy $m_u m_c/m^2_t \ll m_d
m_s/m_b^2$. By writing the most general superpotential and soft terms
one then gets the following textures for quark and squark masses at
the GUT scale:  
\begin{eqnarray}
\label{textures}
M_d &=& m^D
\left(
\begin{array}{ccc}
0 & \frac{\epsilon^{\prime}}{\sqrt{1+\rho^2 k^2}}, & 0 \\
-\epsilon^{\prime} & 0 & \epsilon e^{i \phi} \\
0 & \rho & 1
\end{array} \right) \\
M_u &=& m^U
\left(
\begin{array}{ccc}
0 & c\epsilon\epsilon^{\prime} & 0 \\
-c\epsilon\epsilon^{\prime} & 0 & a\epsilon \\
0 & b\epsilon e^{i \psi} & 1
\end{array} \right)\\
m^2_Q &=& m^2_{3/2}
\left(
\begin{array}{ccc}
1& 0 & \alpha\epsilon\epsilon^{\prime} \\
0 & 1 & 0 \\
\alpha^*\epsilon\epsilon^{\prime} & 0 & r_3
\end{array} \right) \\
m^2_d &=& m^2_{3/2}
\left(
\begin{array}{ccc}
1& 0 & \alpha^{\prime}\epsilon\epsilon^{\prime} \\
0 & 1+ \lambda |\rho|^2 & \beta \rho^* \\
{\alpha^{\prime}}^*\epsilon\epsilon^{\prime} & \beta^* \rho & r^{\prime}_3
\end{array} \right) \\
m^2_u &=& m^2_{3/2} 
\left(
\begin{array}{ccc}
1& 0 & \alpha^{\prime \prime}\epsilon\epsilon^{\prime} \\
0 & 1 & 0 \\
{\alpha^{\prime \prime}}^*\epsilon\epsilon^{\prime} & 0 & r^{\prime \prime}_3
\end{array}\right)\; ,
\end{eqnarray}
where $\epsilon = \ord{V/M}$, $\epsilon^{\prime}=\ord{v/M}$,
$\rho=\ord{V/M^{\prime}}$ and all other coefficients arise from
couplings of order one.  The parameters $r_3$, $r^{\prime}_3$,
$r^{\prime \prime}_3$ differentiate the third sfermion family masses
from the $U(2)$ invariant masses of the first two families. They can
differ from one since the flavour symmetry does not constrain this
ratio. For simplicity, from now on   we will assume $r_3 =
r^{\prime}_3 = r^{\prime \prime}_3$.

Some comments are in order. Since $(M_d)_{22} = 0$, an asymmetry
$(M_d)_{32} > (M_d)_{23}$ is required in order to agree with the
relation $(m_s/m_b)_{\rm GUT} \sim |V_{cb}|_{\rm GUT}$ without
invoking cancellations between the contributions to $V_{cb}$ from
$M_d$ and $M_u$.  Such an asymmetry is obtained here because
$(M_d)_{32}$ is generated by exchange of the $U(2)$ singlets
$H^{\prime}_1$, $H^{\prime}_2$ at the scale $M^{\prime}\sim V$,
whereas $(M_d)_{23}$ is generated by the exchange of the $U(2)$
doublets $\chi^a$, $\bar\chi_a$ at the higher scale $M\gg V$. The same
singlet exchange splits the masses of the first two families in the
down-right sector.  Since the $U(2)$ singlets $H^{\prime}_1$,
$H^{\prime}_2$ are $SU(5)$ singlets, they do not contribute at first
order to the up-quark mass matrix: both $(M_u)_{23}$ and $(M_u)_{32}$
are of order $\epsilon$.  The larger hierarchy $m_c/m_t \ll m_s/m_b$
follows.  As for the further suppression of $m_u m_c/m_t^2$ with
respect to $m_d m_s/m_b^2$, it is due here to the invariance of
$A^{ab}$ under $SU(5)$ \cite{BHRR,BH2}. The operator $A^{ab}T_a T_b
H$, in standard $SU(5)$ notations, does in fact vanish due to the
antisymmetry of $A^{ab}$. $SU(5)$ breaking effects must be included in
order to generate a non-vanishing $(M_u)_{12}$ entry, thus giving the
extra $\epsilon$ there. Finally, the factor $(1+\rho^2 k^2)^{-1/2}$ in
the $(M_d)_{12}$ entry comes from the diagonalization of the kinetic
terms.  Notice that, thanks to rephasing invariance, we have the
freedom to have all real entries apart from $(M_d)_{23}$ and
$(M_u)_{32}$. We choose to work with real parameters, and so
explicitly write these phases in terms of two angles $\phi$ and
$\psi$.

We do not discuss here the A-terms. The flavour symmetry
constrains them to have the same structure of the Yukawa couplings. 
Once the constraints from $\Delta F=1$ processes (and EDMs) have been
taken into account\footnote{Notice that indeed the saturation of 
$\varepsilon^{\prime}/\varepsilon$ can be obtained even for tiny values 
of the corresponding A-parameters~\cite{epss}.},
the contributions to the $\Delta F=2$ transitions
relevant to the UT fit are negligible \cite{constraints}. We can therefore
safely drop these terms in the following. 

One important property of the flavour structure in
eq.~(\ref{textures}) is the presence of a large mixing between the
second and third generation in the right-handed sector. This is
irrelevant for SM contributions to flavour-changing processes, but has
a large impact in the sfermionic sector. Indeed, squark exchange with
this mixing can generate large coefficients for the Left-Right
four-fermion operators in the $\Delta F=2$ effective Hamiltonian,
which are then enhanced both by the QCD running and by the
matrix elements. Therefore, we are in the interesting situation in
which there is a complementary sensitivity of SUSY contributions to
those features of the flavour structure that cannot be probed
considering only SM-induced amplitudes. This explains why in this case
it is very important to include SUSY effects when testing the flavour
structure of the model. The same considerations apply, as we shall see
in the following, to the model based on a $SU(3)$ flavour symmetry. 

\subsection*{Unitarity Triangle Analysis}
As discussed in the Introduction, our aim here is to show how SUSY
effects can modify the predictions of flavour models, and in
particular how the shape of the UT depends on the contributions from
the SUSY sector. In general, some of the parameters of the flavour
model can be determined using only SM-dominated (tree-level)
processes. However, the CP-violating
phases and the sfermion mass parameters can only be extracted from
loop processes. In principle, one should proceed by simultaneously
fitting all these parameters. Unfortunately, at present this is not
possible since the only relevant quantities that have been measured are
$\varepsilon_K$ and  $\Delta m_{B_d}$, together with the lower bound on
$\Delta m_{B_S}$. When, hopefully in the near future, more experimental
data will be available (a more precise measurement of $a_{J/\psi
  K}$, CP-asymmetries in other channels, rare decays, etc.), a global
fit will be feasible. For the purpose of illustrating the potentially
large effects due to SUSY contributions, we can however proceed by
fixing the CP phases in the Yukawa couplings to some representative
values. We then scan over the sfermionic parameter space imposing
 $\varepsilon_K$,  $\Delta m_{B_d}$ and $\Delta m_{B_s}$
constraints and obtain
predictions for other observables as a function of SUSY
parameters. Once new measurements are available, these predictions can
be turned into further constraints on the SUSY parameter space.

For our numerical analysis, we first run with SUSY one-loop
renormalization group equations the mass matrices from the GUT to the
electroweak scale~\cite{Barbieri:1995tw}. We then use the NLO QCD running
\cite{Ciuchini:1998bw,buras} from the electroweak scale to the
hadronic scale for the $\Delta F=2$ amplitudes and take the relevant
$B$-parameters from lattice QCD, whenever they are available.  In
particular, we use the NLO $\Delta S=2$ effective Hamiltonian in the
Landau RI scheme (LRI) as given in ref.~\cite{Ciuchini:1998ix} and the
corresponding $B$ parameters from
ref.~\cite{Allton:1999sm}. Concerning the $\Delta B=2$ $B$-parameters,
only one of the three we need is available at present, and we have
taken it from ref.~\cite{Becirevic:2000nv}.

The first step of the analysis is to fit the parameters entering the
fermionic matrices for fixed values of the phases, to reproduce the
experimental values for fermion masses and $|V_{ub}|$, $|V_{us}|$
and $|V_{cb}|$, which
can be determined using tree-level weak decays. In table~\ref{tab:u2}
we report some numerical examples for different choices of the phase.
The fit uses the
values in table \ref{tab:exp1} as input parameters.

\begin{table}
\begin{center}
\begin{tabular}{||c||c|c|c|c||}
\hline \hline
$\phi$ & 0 & -0.25 & -0.25 & -0.5\cr
$\psi$ & 0 & 0 & -0.25 & -0.25\cr \hline
$\epsilon$ & 0.059 & -0.055 & 0.073  & 0.064 \cr
$\epsilon^{\prime}$ &  0.0064 & -0.0058 & -0.0054 & -0.0065 \cr
$\rho$ &0.49 & 0.49 & -0.33 & -0.46\cr
a & 1.13 & 1.11 & 1.03 & 0.88\cr
b & -3.34 & -3.23 & 1.91 &  -2.46\cr
c & 1.03 & 0.87 & 0.71 & -0.82\cr
k & -0.75 & -0.46 & -1.07 & -0.77\cr
\hline
$\bar \rho$ & 0.428 & 0.357 & 0.253 & 0.246 \cr
$\bar \eta$ & 0 & 0.168 & 0.164 & 0.365 \cr
$\varepsilon^{SM}_K$ &0&0.00103&0.00124& 0.00255\cr
$a_{J/\psi K}^{SM}/\eta_{CP}$ &0&0.489&0.418& 0.784\cr
$a_{J/\psi \phi}^{SM}/\eta_{CP}$ &0&-0.016&-0.017&-0.038\cr
$|\Delta m_{B_b}^{SM}|$ &0.196&0.249&0.358& 0.409\cr
$|\Delta m_{B_s}^{SM}|$ &16.0&16.1&16.3& 15.5 \cr
\hline \hline
\end{tabular}
\caption{\it Results of the fit of fermionic parameters for different
  choices of the phases $\psi$ and $\phi$ (see text for details) in
  the $U(2)$ case. The values in the first half of the table
  correspond to the fitted parameters, and the results in the second
  half correspond to the purely SM contributions to $\Delta F=2$
  processes. The mass differences are given in ps$^{-1}$.
  $\bar{\rho}$ and $\bar{\eta}$ appear in the Wolfenstein
  parameterization of the CKM matrix~\cite{burasbig}.  The definition
  of the asymmetries is according to ref.~\cite{branco}.}
\label{tab:u2} 
\end{center}
\end{table}

\begin{table}
\begin{center}
\begin{tabular}{||c|c|c|c||}
\hline \hline
& Value & Error & Ref.\cr
\hline \hline
$|V_{us}|$ & $0.2237$ & $0.0037$ & \cite{Ciuchini:2000de}\cr
$|V_{ub}|$ & $35.5 \time 10^{-4}$ & $3.6 \times 10^{-4}$ & 
                    \cite{Ciuchini:2000de}\cr
$|V_{cb}|$ & $41.0 \times 10^{-3}$ & $1.6 \times 10^{-3}$ & 
                    \cite{Ciuchini:2000de}\cr
$m_t(m_t)$ & 167  & 5  & \cite{massatop} \cr
$m_c(2 GeV)$ & 1.48  & 0.28  & \cite{Allton:1994ae} \cr
$m_b(M_b)$ & 4.26  & 0.09   & \cite{Gimenez:2000cj}\cr
$m_s(2 GeV)$ & 0.120  &  0.009   & \cite{Becirevic:2000kb}\cr
$Q \equiv \frac{m_s/m_d}{\sqrt{1- (m_u/m_d)^2}}$ & 22.7 
                   &0.8&\cite{Leutwyler:1996ej}\cr
$m_s/m_d$ & 21 & 4 & \cite{Groom:2000in}\cr
\hline
$\tan\beta$ & 3 &&\cr
$\sin^2 \theta_W $& 0.23117 &&\cr
$M_Z$ & $91.188$  &&\cr
$M_{\rm GUT}$ & $2 \times 10^{16}$  &&\cr
$M_g(M_Z)$ & 500  &&\cr
$m_{3/2}$ & 200  &&\cr
$\alpha_{QCD}(M_Z)$ & $0.119$ &&\cr
\hline
$|\varepsilon_K|$& $2.271\times 10^{-3}$&$0.017\times
10^{-3}$&\cite{Groom:2000in}\cr 
$\Delta m_{B_d}$ & 0.487 & 0.014 & \cite{lepbosc}\cr
$\Delta m_{B_s}$ & $>14.5$ (95\% c.l.)&& \cite{lepbosc}\cr
$a_{J/\psi K}/\eta_{CP}$& 0.48 & 0.16& \cite{sin2beta}\cr
\hline
$\Delta m_K$ & $3.495 \times 10^{-15}$  & $0.013\times 10^{-15}$ &\cr
$m_{B_d}$ & 5.279  & 0.002  &\cr
$m_{B_s}$ & 5.369  & 0.002  &\cr
$m_{K^0}$ &  0.497672  & 0.000031  &\cr
$f_{B_d}$ & $0.174$  &  $0.022$ & \cite{Becirevic:2000nv}\cr
$f_{B_s}$ & $0.204$  &  $0.015$ & \cite{Becirevic:2000nv}\cr
$f_K$ & $0.161$  & $0.0015$  &\cr
$\hat{B}_{B_d}^{Q_1}$ & $1.38$ & $0.11$ & \cite{Becirevic:2000nv}\cr
$\hat{B}_{B_s}^{Q_1}$ & $1.35$ & $0.05$ & \cite{Becirevic:2000nv}\cr
$B_K^{\bar{\scriptscriptstyle{MS}}}(2 GeV)_{Q_1}$ & $0.61$ & $0.06$ &
\cite{Ciuchini:1998ix}\cr 
$B_K^{LRI}(2 GeV)_{Q_4}$ & $1.04$ & $0.06$ &\cite{Ciuchini:1998ix}\cr
$B_K^{LRI}(2 GeV)_{Q_5}$ & $0.73$ & $0.10$ &\cite{Ciuchini:1998ix}\cr
\hline \hline
\end{tabular}
\caption{\it Experimental data and fixed parameters in the
  analysis. The $B$ mass differences are given in ps$^{-1}$, the $K$
  mass difference and all other masses in GeV. $M_g(M_Z)$ is the
  gluino mass at the electroweak scale and $m_{3/2}$ is the
  mass of the first two generations of sfermions at the GUT scale.
  $a_{J/\psi K}/\eta_{CP}$ is the world average of asymmetry
  measurements (normalized for CP-even final states).
  $\hat{B}_{B_d}^{Q_1}$ and $\hat{B}_{B_s}^{Q_1}$ are the
  renormalization group invariant B-parameters for the SM $\Delta B=2$
  operators. $B_K^{\bar{\scriptscriptstyle{MS}}}(2 GeV)_{Q_1}$ is the
  B-parameter in the $\bar{MS}$ scheme for the SM $\Delta S=2$
  operator, and $B_K^{LRI}(2 GeV)_{Q_{4,5}}$ are the B-parameters in
  the Landau RI scheme for the SUSY $\Delta S=2$ operators $Q_{4,5}$
  (see ref.~\cite{Ciuchini:1998ix} for details).}
\label{tab:exp1}
\end{center}
\end{table}

\begin{figure}
\begin{center}  
\epsfysize=7cm 
\epsfxsize=11cm 
\epsffile{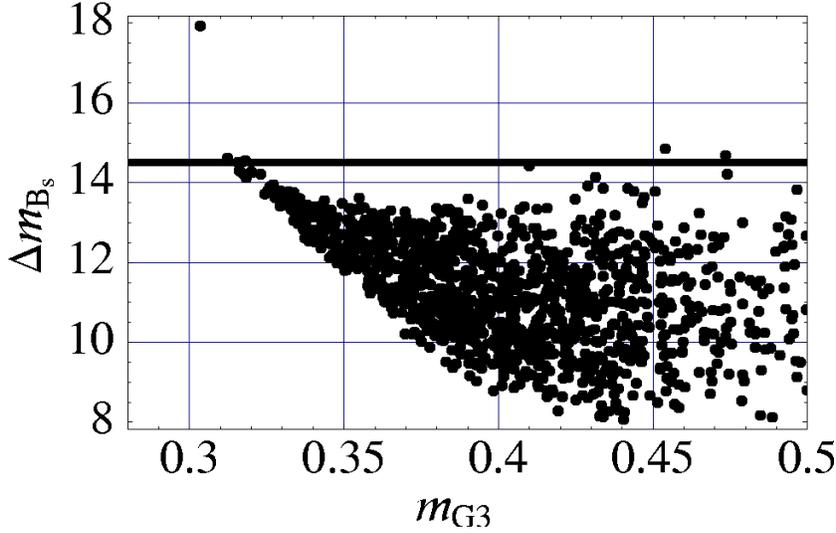}
\end{center}
\caption{\it Dependence of $\Delta m_{B_s}$ (in ps$^{-1}$) on $m_{G3}$ 
  (in TeV), the GUT scale mass of the third family. Here the Yukawa
  couplings are assumed to be real ($\phi=\psi=0$). The line
  represents the lower bound from experiments $\Delta m_{B_s}>14.5$
  ps$^{-1}$~\cite{lepbosc}.}
\label{fig:dmsu2no}
\end{figure}

The second step is to constrain the SUSY parameters making use of
$\varepsilon_K$ and $\Delta m_{B_d}$.~\footnote{For our choice of SUSY
  parameters the gluino exchange represents the dominant SUSY
  contribution. We performed the actual computation of $\Delta F=2$
  amplitudes in the mass insertion approximation
  (MIA)~\cite{constraints}.  Given the particular textures we are
  using for sfermionic soft mass terms, to obtain a reliable result in
  MIA for $\Delta S=2$ observables, multiple mass insertions have been
  included.} We can then predict $\Delta m_{B_s}$, $a_{J/\psi K}$ and
$a_{J/\psi \phi}$ for each given set of SUSY masses compatible with
the constraints. First of all, we note that for vanishing phases in
the Yukawa couplings, once the $\varepsilon_K$ and $\Delta m_{B_d}$
constraints are imposed, the predicted value of $\Delta m_{B_s}$ is
below the present lower bound for almost any choice of SUSY
parameters. The reason for this is the following. For vanishing CKM
phase, the UT collapses to the positive $\bar \rho$ axis, which
implies that the SM contribution to $\Delta m_{B_d}$ is about one half
of the experimental value. While this can be compensated by a large
SUSY contribution, the flavour structure then forces the SUSY
contribution to $\Delta m_{B_s}$ to interfere destructively with the
SM one, resulting inevitably in a too low value for the $B_s - \bar
B_s$ mass difference (see fig.~\ref{fig:dmsu2no}).
\begin{figure}
\begin{center}  
\epsfysize=7cm 
\epsfxsize=11cm 
\epsffile{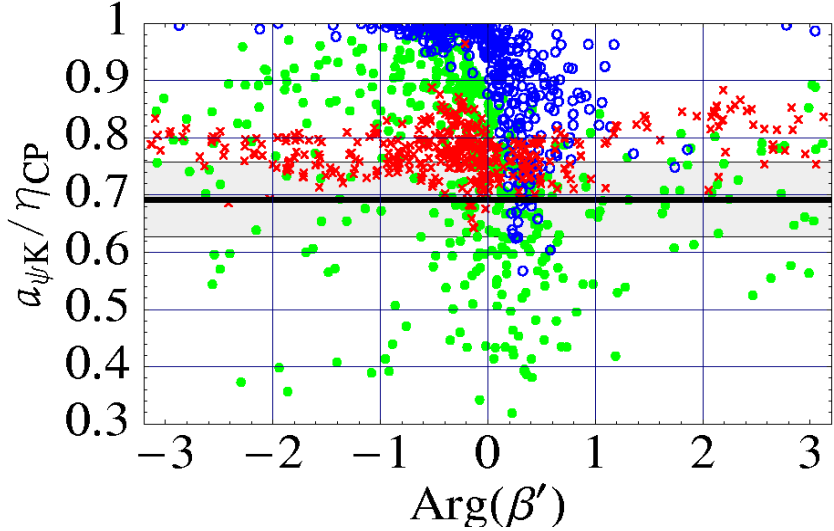}
\end{center}  
\caption{\it Dependence of the time dependent CP asymmetries in $B_d$
  system on the phase of $\beta^{\prime}$, for ($\phi=-0.25,\psi=0$)
  ($\circ$), ($\phi=-0.25,\psi=-0.25$) ($\bullet$) and
  ($\phi=-0.5,\psi=-0.25$) ($\times$).  The thick line with the
  shadowed region corresponds to the SM prediction $a_{J/\psi
    K}/\eta_{CP} = 0.692 \pm 0.065$ \cite{Ciuchini:2000xh}.}
\label{fig:aku2}
\end{figure}
\begin{figure}
\begin{center}  
\epsfysize=7cm 
\epsfxsize=11cm 
\epsffile{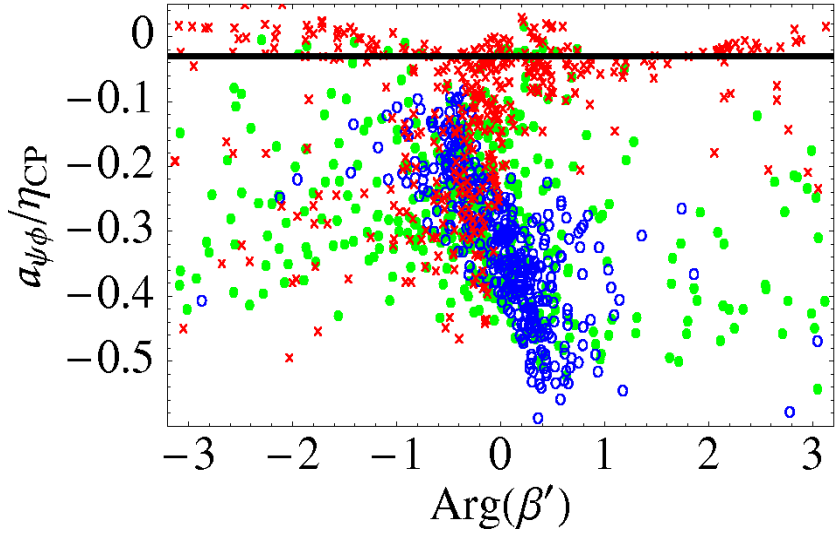}
\end{center}  
\caption{\it Dependence of the time dependent CP asymmetries in $B_s$
  system on the phase of $\beta^{\prime}$, for ($\phi=-0.25,\psi=0$)
  ($\circ$), ($\phi=-0.25,\psi=-0.25$) ($\bullet$) and
  ($\phi=-0.5,\psi=-0.25$) ($\times$).  The thick line is the SM
  prediction $\sim -3$\% \cite{Hurth:2001yy}.}
\label{fig:aphiu2}
\end{figure}
\begin{figure}
\begin{center}  
\epsfysize=7cm 
\epsfxsize=11cm 
\epsffile{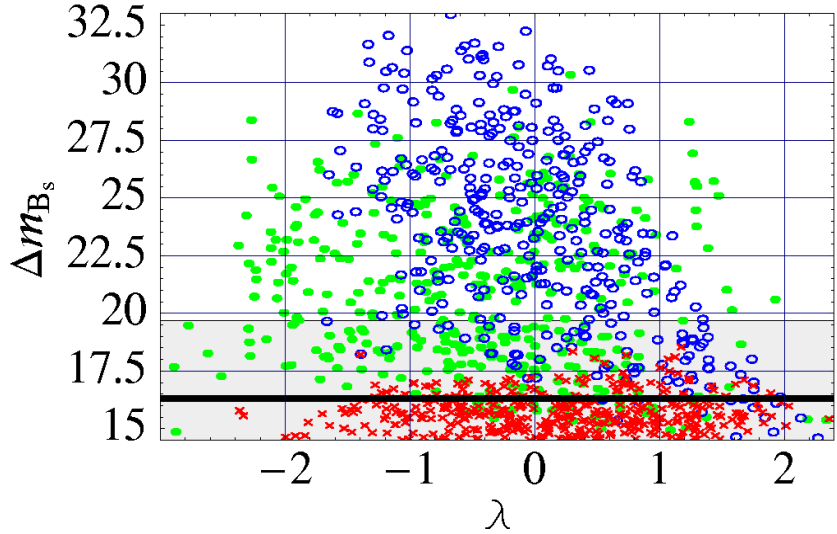}
\end{center}  
\caption{\it Dependence of $\Delta m_{B_s}$ (in ps$^{-1}$)  on $\lambda$, 
 for ($\phi=-0.25,\psi=0$) ($\circ$), ($\phi=-0.25,\psi=-0.25$) 
($\bullet$) and ($\phi=-0.5,\psi=-0.25$) ($\times$). 
The thick line with the shadowed region is the SM prediction
$\Delta m_{B_s}=16.3 \pm 3.4$ ps$^{-1}$ \cite{Ciuchini:2000de}.}
\label{fig:dmsu2}
\end{figure}

Once we introduce CP violation in the CKM matrix this anticorrelation
between SUSY contributions to $\Delta m_{B_d}$ and $\Delta m_{B_s}$ is lost,
and good fits can be obtained also for relatively small values of the
CKM phase. This is interesting since, as we anticipated in the
introduction, not only can we successfully reproduce all the observed
CP violation, but thanks to SUSY contributions we can obtain values
for $\Delta m_{B_s}$, $a_{J/\psi K}$ and $a_{J/\psi \phi}$ that can
considerably differ from the SM predictions. As an example, we report
in figs.~\ref{fig:aku2}, ~\ref{fig:aphiu2} and~\ref{fig:dmsu2}, the
scatter plots for $\Delta m_{B_s}$, $a_{J/\psi K}$ and $a_{J/\psi
  \phi}$ for  non-vanishing CKM phases, to be compared with the
predictions of the standard UT analysis (see for example
ref.~\cite{Ciuchini:2000de} for up-to-date results) and the SM
prediction $a_{J/\psi \phi} \simeq 0$. Notice that, as expected,
for increasing phases $\phi$ and $\psi$ the prediction 
tends to reproduce the SM ones,
due to the fact that SUSY is playing a weaker role. Indeed,
it is possible to show that this model can reproduce $\varepsilon_K$
and $\Delta m_{B_d}$
also with vanishing SUSY contributions~\cite{toappear}.

\section{A model with an SU(3) flavour symmetry}

In this case quark Superfields are assigned to transform as a triplet 
under $SU(3)$ to be denoted by $\psi_i \sim 3$.
This model is very similar to the one discussed in~\cite{rossi2}.
The flavons in the model are $S^{ij} \sim \bar{6}$ and $\phi_i \sim 3$.
Another singlet $T^i_j \sim 8$, not directly coupled to matter Superfields,
 is required to get phenomenologically acceptable textures (it is responsible 
for the appearance of the parameter $b$, see below).
The breaking pattern associated to $SU(3)$ breaking fields directly coupled 
to SM fermions\footnote{The auxiliary fields in~\cite{rossi2} modify the
 breaking pattern but as far as the observable sector is concerned 
the effective breaking is the one shown.} is
\[
SU(3)\stackrel{<S^{33}>}{\longrightarrow}
SU(2)\stackrel{<\phi>}{\longrightarrow}
\emptyset.
\]
The symmetry violating operators involving the lighter families
are suppressed by the flavons VEVs over the scale of symmetry
breaking messengers in the FN mechanism. The suppression
factors we get are $1 > \eta > \epsilon > \epsilon^{\prime}$ 
in the equations below.

The textures we get are (neglecting higher order terms)
\begin{eqnarray}
\label{down}
M_d&=&m^D  \left(\begin{array}{ccc} 0 & \epsilon^{\prime} & 0 \\
       - \epsilon^{\prime} & c \eta                 & b \epsilon \\    
       0                   & \epsilon          & \eta  
\end{array}\right)\\
\label{fermionic}
M_d&=&m^U  \left(\begin{array}{ccc} 0 & 0 & 0 \\
       0 &    c \eta         & 0 \\    
       0     & 0         & \eta  
\end{array}\right)\\   
m^2_Q &=& m^2_{3/2}
\left(
\begin{array}{ccc}
1& 0 & \alpha\epsilon\epsilon^{\prime} \\
0 & 1+\lambda \epsilon^2 & \beta \epsilon \eta \\
\alpha^*\epsilon\epsilon^{\prime} & \beta^*\epsilon\eta & r_3
\end{array} \right) \\
m^2_d &=& m^2_{3/2}
\left(
\begin{array}{ccc}
1& 0 & \alpha^{\prime}\epsilon\epsilon^{\prime} \\
0 & 1+ \lambda^{\prime} \epsilon^2  & \beta^{\prime} \epsilon \eta \\
{\alpha^{\prime}}^*\epsilon\epsilon^{\prime} & {\beta^{\prime}}^{ \ast} \epsilon \eta & r_3
\end{array} \right) \\
m^2_u &=& m^2_{3/2} 
\left(
\begin{array}{ccc}
1 & 0 & 0 \\
0 & 1 & 0 \\
0 & 0 & r_3 
\end{array}\right)\;.
\end{eqnarray}
where $c \simeq m_c/m_t$, $m^U$ and  $m^D$ are proportional to the masses of
top and bottom quark respectively.
As in the $U(2)$ case, $r_3$ denotes the ratio
$m^2_{G_3}/m^2_{3/2}$. Although for unbroken $SU(3)$ one has $r_3=1$,
the large breaking can generate a mass splitting between the third and
first two generations of order one.   

Comparing the Yukawa couplings to the ones in ref.~\cite{rossi2}, one
sees that the (1,3) and (3,1) entries are missing in our case. This
implies that the CKM phase in the present model is negligibly small
(proportional to $m_c/m_t$). However, as we shall see in the
following, we are able to explain $\varepsilon_K$ with SUSY
contributions and fit the UT, and therefore we do not need to
introduce these additional entries.  Notice that the reality of the
fermionic mass matrices is not another assumption added by hand, but just a
consequence of the structure of the textures, that always allows to
redefine the fermionic fields in such a way as to remove all the complex
phases (an explicit check of this property can be achieved with the
Jarlskog determinant~\cite{Jarlskog}).  The $U(2)$ model
presented in the previous section does not share this property, due to
the non-trivial structure of the up-type quark mass matrix, and indeed
the fit of the model required a sizable complex phase in the CKM
matrix, as discussed before. The possibility of fitting all CP
violating observables with a real CKM matrix is indeed an interesting
property of this $SU(3)$ model. 

Just as in the case of $U(2)$, we have a large mixing between second
and third generation in the right-handed sector, due to the presence
of the asymmetry parameter $b$. Therefore, also in this case one can
have large SUSY contributions to $\Delta F=2$ processes
induced by sfermion mixing in the right-handed sector (see the
discussion below eq.~(\ref{textures})).

\subsection*{Unitarity Triangle Analysis}
Since in this case we can neglect the CKM phase, we can separately fit
the Yukawa couplings to the SM-dominated quantities, and the SUSY
parameters to $\Delta F=2$ amplitudes. In this case, the UT collapses
to a line, but in the region of negative $\bar \rho$. This means that
the SM contribution to $\Delta m_{B_d}$ is exceedingly large ($1.04$
ps$^{-1}$).  This is compensated by SUSY contributions. The predicted
amplitude for $\Delta m_{B_s}$ can be much larger than given by the
standard UT analysis, and the CP asymmetries $a_{J/\psi K}$ and
$a_{J/\psi \phi}$ can also differ in sizable way from the SM
prediction.

\begin{table}
\begin{center}
\begin{tabular}{||c|c||}
\hline \hline
$\epsilon$ & -0.31\cr
$\epsilon^{\prime}$ & -0.0053\cr
b &0.10\cr
\hline
$\bar \rho$ &-0.35\cr
$\bar \eta$ & 0\cr
$\varepsilon^{SM}_K$ &0\cr
$a_{J/\psi K}^{SM}/\eta_{CP}$ &0\cr
$a_{J/\psi \eta}^{SM}/\eta_{CP}$ &0\cr
$|\Delta m_{B_b}^{SM}|$ &1.04\cr
$|\Delta m_{B_s}^{SM}|$ &14.0 \cr
\hline \hline
\end{tabular}
\caption{\it Results of the fit of fermionic parameters for different
  in $SU(3)$ with real CKM in. The values in the first half of the table
  correspond to the fitted parameters, and the results in the second
  half correspond to the purely SM contributions to $\Delta F=2$
  processes. The mass differences are given in ps$^{-1}$.}
\label{tab:fitu3} 
\end{center}
\end{table}

In table~\ref{tab:fitu3} we report the fitted values of the fermionic
parameters and the purely SM contributions to $\Delta F=2$
processes.
The parameter $b$, responsible for the large asymmetry between
the entries $M^D_{23}$ and $M^D_{32}$, is generated, as
explained in detail in ref.~\cite{rossi2}, by a $SU(3)$ breaking in
the adjoint representation, which is however not directly coupled to
matter fields. 
We assume $SU(3)$ breaking to take place at a scale 
near the fundamental one, and we take $\eta = 0.7$, compatibly
with this assumption.

We notice that all the solutions we find correspond to relatively
small phases also in the SUSY sector. One may then think that this
model could be embedded in some ``approximate CP''
scenario~\cite{nir}. 

\begin{figure}
\begin{center}  
\epsfysize=7cm 
\epsfxsize=11cm 
\epsffile{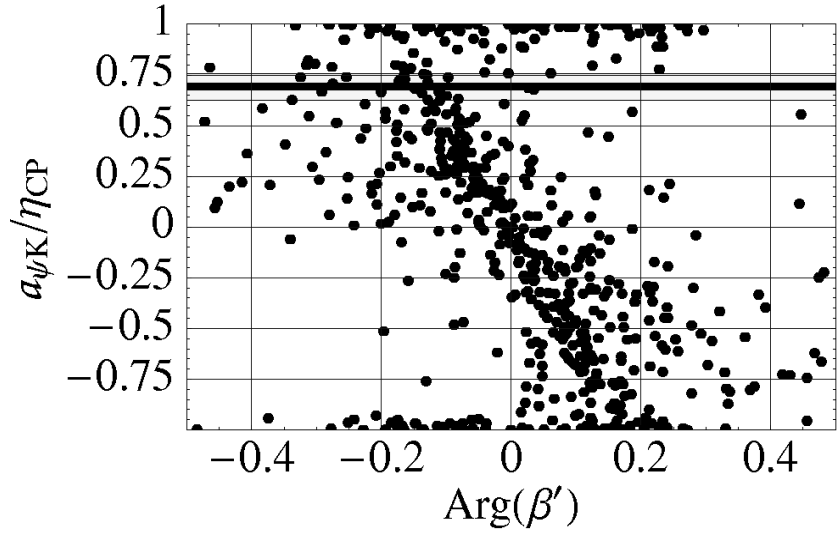}
\end{center}  
\caption{\it Dependence of the time dependent CP asymmetries in $B_d$
  system on the phase of $\beta^{\prime}$ in $SU(3)$ with real CKM.
  The thick line with the shadowed region corresponds to the SM
  prediction $a_{J/\psi K}/\eta_{CP} = 0.692 \pm 0.065$
  \cite{Ciuchini:2000xh}.}
\label{fig:aksu3}
\end{figure}

\begin{figure}
\begin{center}  
\epsfysize=7cm 
\epsfxsize=11cm 
\epsffile{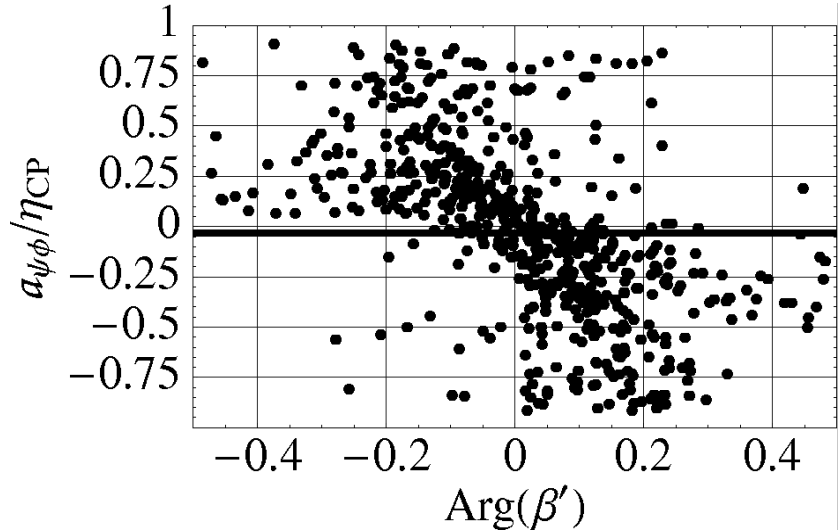}
\end{center}  
\caption{\it Dependence of the time dependent CP asymmetries in $B_s$
system on the phase of $\beta^{\prime}$ in $SU(3)$ with real CKM.
The thick line is the SM prediction $\sim -3$\% \cite{Hurth:2001yy}.}
\label{fig:aphisu3}
\end{figure}

\begin{figure}
\begin{center}  
\epsfysize=7cm 
\epsfxsize=11cm 
\epsffile{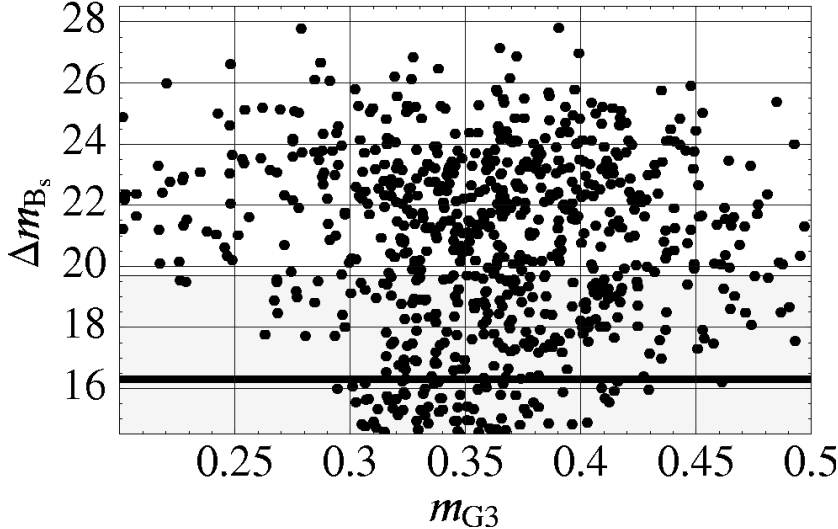}
\end{center}  
\caption{\it Dependence of $\Delta m_{B_s}$ (in ps$^{-1}$) on $m_{G3}$
  (in TeV), the GUT scale mass of the third generation squarks, in
  $SU(3)$ with real CKM.  The thick line with the shadowed region
  corresponds to the SM prediction $\Delta m_{B_s}=16.3 \pm 3.4$
  ps$^{-1}$\cite{Ciuchini:2000de}.}
\label{fig:dmssu3}
\end{figure}

For illustrative purposes, we report in figures~\ref{fig:aksu3},
\ref{fig:aphisu3} and \ref{fig:dmssu3} the scatter plots for the
$a_{J/\psi K}$ and $a_{J/\psi \phi}$ asymmetries and for $\Delta
m_{B_s}$. Similar plots can be obtained as a function of the other
parameters. We see that large values of both $a_{J/\psi \phi}$ and  $\Delta
m_{B_s}$ can be obtained, which would unambiguously signal new physics
contributions. Also small values of $a_{J/\psi K}$ are possible. 


\section{Conclusions}

We have studied SUSY virtual effects in two nonabelian flavour
models, in which both the flavour structure of the fermionic and the
sfermionic sectors are tightly constrained. We have explicitly shown
the relevance of SUSY corrections, and discussed how these may modify
the UT fit in these models and  generate significant deviations from SM
predictions for three theoretically clean observables: $a_{J/\psi K}$,
$a_{J/\psi \phi}$ and $\Delta m_{B_s}$. In the model based on a $U(2)$
flavour symmetry, where CP violation is present in the CKM matrix and
a good fit can also be obtained in the limit of negligible SUSY
contributions, the shape of the UT can be sizably modified for SUSY masses
around $500$ GeV, resulting in large values of the CP asymmetry in
$B_s \to J/\psi \phi$ decays and of $\Delta m_{B_s}$.
In the $SU(3)$ model, the CKM matrix is real to a very good
approximation, and the UT collapses to a line with negative $\bar
\rho$; however, for SUSY masses around $500$ GeV, sparticle
contributions can account for all of the observed CP violation, while
large deviations from the SM predictions for $a_{J/\psi K}$,
$a_{J/\psi \phi}$ and $\Delta m_{B_s}$ are possible. 
 
In conclusion, the role played by SUSY in FCNC and CP violating 
processes crucially depends on the nature of the mechanism
which originates the SUSY breaking and transmits the information to the 
observable sector. A first, plausible option is that such mechanism
has nothing to do with what gives rise to the flavour
structure of the theory. The MFV situation is encountered in classes of
SUSY models: anomaly, gauge, gaugino mediated SUSY breaking mechanisms
constitute interesting examples.
In these cases the hopes to indirectly observe SUSY manifestations in FCNC
are rather slim: the CP asymmetry in $b \rightarrow s \gamma$ or the $\gamma$
angle of the UT are certainly interesting possibilities, but overall 
the general impression is that we will have to wait for direct
detection to have a SUSY signal.
On the contrary, if one turns to gravity mediated SUSY breaking, there
is no particular reason for such flavour blindness. As soon as a new
flavour structure arises in the sfermionic sector, SUSY allows
for quite conspicuous new contributions to FCNC, which in general
are even too large for the tight FCNC experimental constraints.
Among the adopted solutions to this flavour problem, the presence
of an additional non-abelian flavour symmetry stands up as one of
the most attractive possibilities.
In this context, our analysis has considered a couple of interesting
examples. The message which emerges from them is twofold.
On one hand it appears that SUSY plays a major role in the  fit
of the UT. On the other hand it emerges that SUSY flavour models
have concrete potentialities to exhibit sizable departures from the SM 
in particularly clean B-physics observables, while keeping
under control all the other dangerous FCNC threats.
Here the ``competition'' between direct and indirect searches
to give the first hint for SUSY remains still open.

\section*{Acknowledgments}
This work is partially supported by the RTN European Program
HPRN-CT-2000-00148.
We thank O. Vives for stimulating discussions.

\end{document}